\documentclass[showpacs, pra,twocolumn,preprintnumbers ,amsmath, amssymb, superscriptaddress, aps]{revtex4-1}
\usepackage{color}
\usepackage{amsmath,amssymb}
\usepackage{pifont}
\usepackage{amssymb}  
\usepackage{bbold}
\usepackage{float}
\usepackage{subfloat} 

\usepackage[caption=false]{subfig}
\usepackage{tikz}
\usepackage{makecell}
\usepackage{subfig}
\usepackage{pifont}   
\usepackage{graphicx} 
\graphicspath{{Figures/}}
\usepackage{dcolumn}  
\usepackage{bm}       
\usepackage{multirow} 
\usepackage{placeins}
\usepackage[colorlinks]{hyperref}
\usepackage{mathtools}

\newcommand*{\rom}[1]{\expandafter\@slowromancap\romannumeral #1@}
\begin{document}

        \title{Klein tunneling through triple barrier in 
        	AB bilayer graphene}
        \date{\today}
        \author{  Mouhamadou Hassane Saley }
        \email{hassmosy@gmail.com}
        \affiliation{Laboratory of Theoretical Physics, Faculty of Sciences, Choua\"ib Doukkali University, PO Box 20, 24000 El Jadida, Morocco}
        \author{Abderrahim El Mouhafid  }
        \email{elmouhafid.a@ucd.ac.ma}
        \affiliation{Laboratory of Theoretical Physics, Faculty of Sciences, Choua\"ib Doukkali University, PO Box 20, 24000 El Jadida, Morocco}
        \author{Ahmed Jellal}
        \email{a.jellal@ucd.ac.ma}
        \affiliation{Laboratory of Theoretical Physics, Faculty of Sciences, Choua\"ib Doukkali University, PO Box 20, 24000 El Jadida, Morocco}
        \affiliation{Canadian Quantum Research Center, 204-3002 32 Ave Vernon,  BC V1T 2L7, Canada}
          \author{Ahmed Siari}
          \affiliation{Laboratory of Measurement and Control Instrumentation, Faculty of Sciences, Choua\"ib Doukkali University, PO Box 20, 24000 El Jadida, Morocco}           

        
\begin{abstract}                                       
We investigate the transport properties of charge carriers in AB bilayer graphene through a triple electrostatic barrier. We calculate the transmission and reflection using the continuity conditions at the interfaces of the triple barrier together with the transfer matrix method. First, we consider the case where the energy is less than the interlayer coupling $\gamma_1$ and show that, at normal incidence, transmission is completely suppressed in the gap for a large barrier width while it appears in the gap for a small barrier width. For energies greater than $\gamma_1$, we show that in the absence of an interlayer potential difference, transmission is less than that of a single barrier, but in its presence, transmission in the gap region is suppressed, as opposed to a double barrier. It is found that one, two, or three gaps can be created depending on the number of interlayer potential differences applied. Resonance in the $T_-^+$ transmission channel is observed that is not seen in the single and double barrier cases. Finally, we compute the conductance and show that the number of peaks is greater than the double barrier case.
                        
\end{abstract}          
        
\maketitle

        \section{Introduction}
Graphene \cite{graphene}, a gapless semiconductor with just one layer of carbon atoms, was discovered in 2004 by Geim and  Novoselov and is now one of the most active research topics in the field of condensed matter because of its interesting physical properties. In fact, in addition to holding the record for the highest thermal conductivity, graphene presents high electronic quality with mobility of the order of 250 000 $\text{cm}^{2}/\text{Vs}$, 20 times higher than that of GaAs \cite{Morozov,Geim}. With a Young's modulus of $\sim$ 1 TPa and an intrinsic tensile strength of $\sim$ 130 GPa \cite{Changgu}, it is one of the strongest materials discovered.
From the optical point of view, graphene is transparent, it observes only 2.3\% of visible light \cite{Nair}. Due to the symmetry of the graphene crystal structure, its charge carriers, which behave like relativistic particles, are described by the Dirac-like Hamiltonian for massless fermions \cite{Katsnelson_2006,Ramezani,FONSECA20104359}. Graphene also has unusual properties such as Klein tunneling \cite{Katsnelson_2006,Young_2009,Stander} and a non-conventional quantum Hall effect \cite{Novoselov_2005,Zhang,Gusynin}. All of these outstanding features demonstrate graphene's relevance. However, the Klein tunneling is an obstacle in terms of possible applications. A graphene-based transistor, for example, cannot have an on/off switching state, according to \cite{Tudorovskiy_2012,Yung_2013,Ramezani}. It is then crucial to open a gap in the energy spectrum of graphene for its application in semiconducting devices. Fortunately, there are several methods to achieve this goal, such as coupling to a substrate or doping with impurities \cite{Zhou,Jung_2015,Filho,San,Zarenia}.

In the case of graphene bilayer AB (Bernal \cite{Bernal}), a system of two layers of carbon atoms placed on top of each other, it is easy to open and control a gap by applying an external electric field \cite{McCanne,Falko,Barbier}. In contrast to single layer graphene, which has a linear energy band, bilayer graphene has a parabolic energy spectrum with four bands. Furthermore, unlike single-layer graphene, tunneling through a potential barrier at normal incidence is characterized by a perfect reflection \cite{Katsnelson_2006, Tudorovskiy_2012}, anti-Klein tunneling, due to confined state cloaking \cite{GU}. However, in cases of double barrier \cite{Mouhafid,Lu_2015,Bai} the anti-Klein tunneling is destroyed and transmission resonances appear as a consequence of the bound electron states present in the well. In addition, it is shown that in the presence of a single magnetic barrier, the cloak effect is attenuated, so the anti-tunneling is reversed \cite{GU,Agrawal_Garg__2012}. Indeed, several studies on the transport properties of graphene bilayers have been conducted \cite{Katsnelson_2006, Tudorovskiy_2012, Barbier, Ramezani,Snymam, Barbier_2010}, but only for energies less than the interlayer coupling $\gamma_{1} \simeq0.4$ eV \cite{McCanne}. In this case ($E<\gamma_{1}$) only one mode of propagation is available, and then there is only one transmission channel. Recently, some studies \cite{Van,Mouhafid,JELLAL2015149,Redouani,BENLAKHOUY2021114835}, dealing with tunneling through electrostatic and even magnetic barriers, have also taken into account the case of energy higher than the interlayer coupling where two propagation modes exist. As a result, four transmission channels are accessible.

Based on previous works, especially \cite{Mouhafid}, we investigate the transport properties of Dirac fermions in bilayer graphene through a triple barrier structure. Using the transfer matrix approach, we determine the transmission and reflection probabilities as well as the associated conductance. We first evaluate the transmission probability for energy less than $\gamma_{1}$ where we find the dependence of the gap on the barrier widths. On the other hand, four transmission and reflection channels are available for energy greater than $\gamma_1$.  
We show  that the transmission is suppressed in the gap region and the number of gaps created depends strongly on the number of interlayer potential differences applied in each region. In addition, we find resonances that are not found in the case of single \cite{Van} and double barrier \cite{Mouhafid}, especially in the $T_{-}^{+}$ transmission channel. Moreover, the conductance displays more peaks than that in \cite{Mouhafid}.

This paper is organized as follows. In section \ref{TTMM}, we present our system's Hamiltonian and derive the corresponding energy eigenvalues and eigenvectors. In section \ref{Trans}, we use the transfer matrix approach and continuity conditions to establish the transmission and reflection probabilities as well as the related conductance. In section \ref{RRDD}, we present and analyze the numerical results obtained. Finally, we present a summary of the main findings in section \ref{CC}.
        \section{Theory and methods}\label{TTMM}
      \begin{figure}[H]
        \centering
        \includegraphics[width=0.47\textwidth]{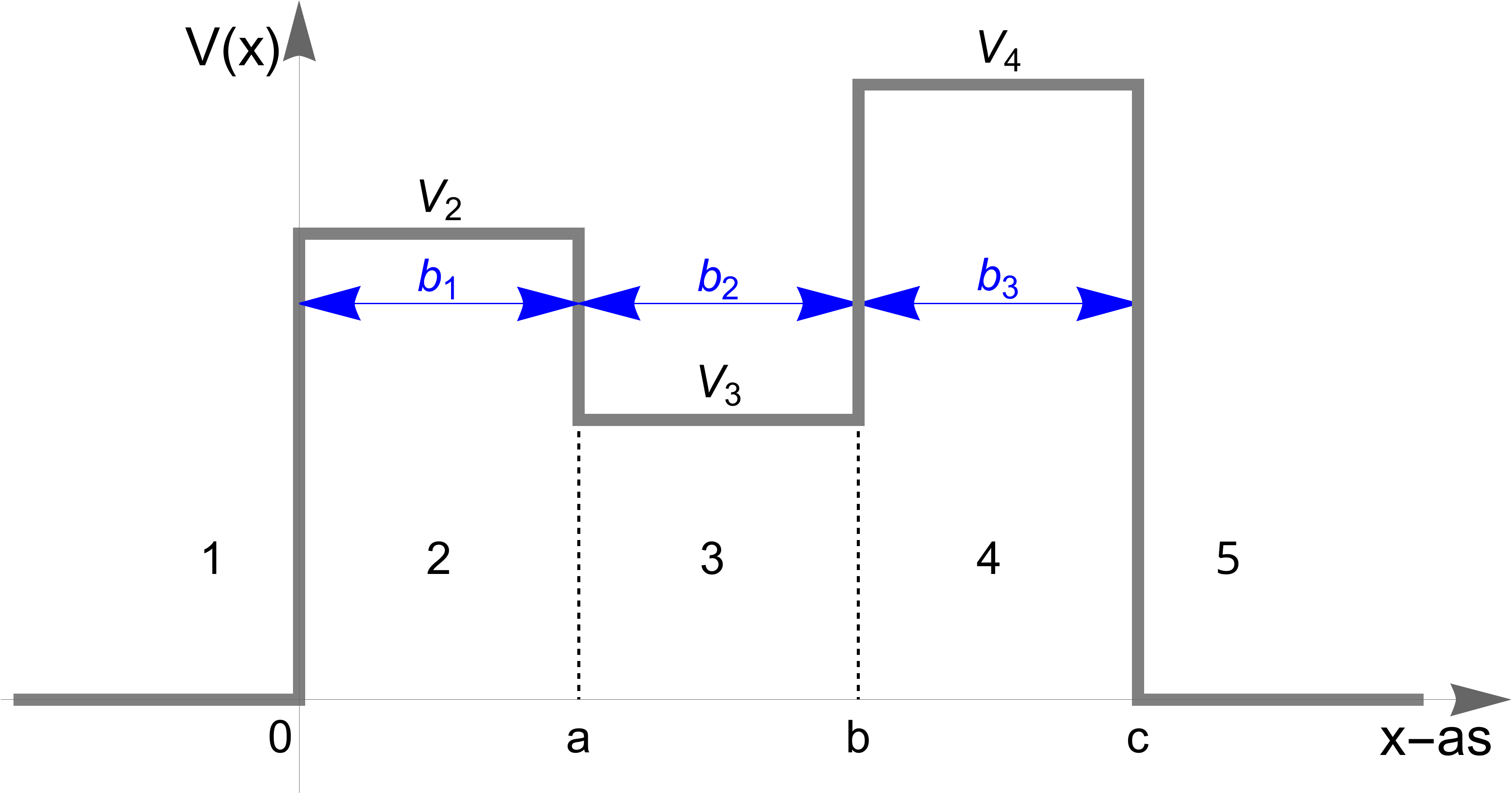}
        \caption{A schematic representation of the triple barrier potential.}
        \label{fig:png2pdf}
\end{figure}
        
We consider AB-stacked bilayer graphene in the presence of triple barrier potential along the $ x $-direction as shown in Fig. \ref{fig:png2pdf}. The Hamiltonian which describes the Dirac fermions in bilayer graphene of AB-stacking is obtained by a nearest-neighbor tight-binding approximation near the $ K $ point \cite{Van,Mouhafid}, which is
\begin{equation}
                H=\begin{pmatrix}
                        V^{+} & v_{F} \pi^{\dagger} &-v_{4} \pi^{\dagger}  & v_{3} \pi \\
                        v_{F} \pi &     V^{+} & \gamma_{1} & -v_{4} \pi^{\dagger}  \\
                        -v_{4} \pi & \gamma_{1} & V^{-} & v_{F} \pi^{\dagger} \\
                        v_{3} \pi^{\dagger} & -v_{4} \pi  &     v_{F} \pi &V^{-}
                \end{pmatrix}
                \label{E1}
\end{equation}
where $\pi=p_{x}+ip_{y}$ denotes the in-plan momenta with $p_{x,y} =-i\hbar \partial _{x,y}$ and  the Fermi velocity is $v_{F}\approx10^{6}$ m/s. The interlayer coupling term is  $v_{3,4}=\frac{v_{F}\gamma_{3,4} }{\gamma_{0}}$, which refers to the coupling between the layers. As shown below, the potentials $ V^{+}$ and $V^{-}$ on the first and second layers are defined in each region denoted by $j$ 
\begin{equation}
        V^{\pm}_{j}=\begin{cases}0, &\qquad j=\text{1, 5}\\
                U_{j}\pm\zeta \delta_{j}, &\qquad j=\text{2, 3, 4}\\
        \end{cases} 
        \label{E2}
\end{equation}
where $U_j$ is the barrier strength, $\delta_j$ is the electrostatic potential difference between layers, and $\zeta=+1$ for the first layer and $\zeta=-1$ for the second layer. 
In the following calculations, we ignore the skew parameters $\gamma_3=0.315$ eV and $\gamma_4=0.044$ eV because they have no effect on the band structure at high energy or transmission at low energy \cite{McCanne, Van}. Therefore, in the $ j $-th region, the Hamiltonian (\ref{E1}) can be written as
\begin{equation}
        H_{j}=\begin{pmatrix}
                U_{j}+\delta_{j} & v_{F} \pi^{\dagger} &0  &0 \\
                v_{F} \pi &     U_{j}+\delta_{j} & \gamma_{1} & 0  \\
                0 & \gamma_{1} &        U_{j}-\delta_{j} & v_{F} \pi^{\dagger} \\
        0 & 0  &        v_{F} \pi &     U_{j}-\delta_{j}
        \end{pmatrix} 
        \label{E3}
\end{equation}
in the basis of  four-component spinor of the form $\psi^{j}(x,y)=[\psi^{j}_{A_{1}},\psi^{j}_{B_{1}},\psi^{j}_{A_{2}},\psi^{j}_{B_{2}}]^{\dagger}$, the symbol $\dagger$ denotes the transpose row vector, with 1 and 2 indexing the two layers. 
Since the momentum is a conserved quantity along the $ y $-direction, then the spinor can be separated as follows
\begin{equation}
\psi^{j}(x,y)=e^{ik_{y}y}[\phi^{j}_{A_{1}},\phi^{j}_{B_{1}},\phi^{j}_{A_{2}},\phi^{j}_{B_{2}}]^{\dagger}.
\label{E4}
\end{equation}
 For simplicity, let us introduce a length scale defined by $l=\frac{\hbar v_{F}}{\gamma_{1}}\approx$1.64 nm. As a result, we can define dimensionless quantities that are $E_{j}\equiv \frac{E_j}{\gamma_{1}}$, $U_{j}\equiv\frac{U_{j}}{\gamma_{1}}$, $\delta_{j}\equiv\frac{\delta_{j}}{\gamma_{1}}$, $x\equiv\frac{x}{l}$, and $k_y\equiv lk_{y}$.
 
\begin{figure}
	\centering
	\includegraphics[width=0.5\textwidth]{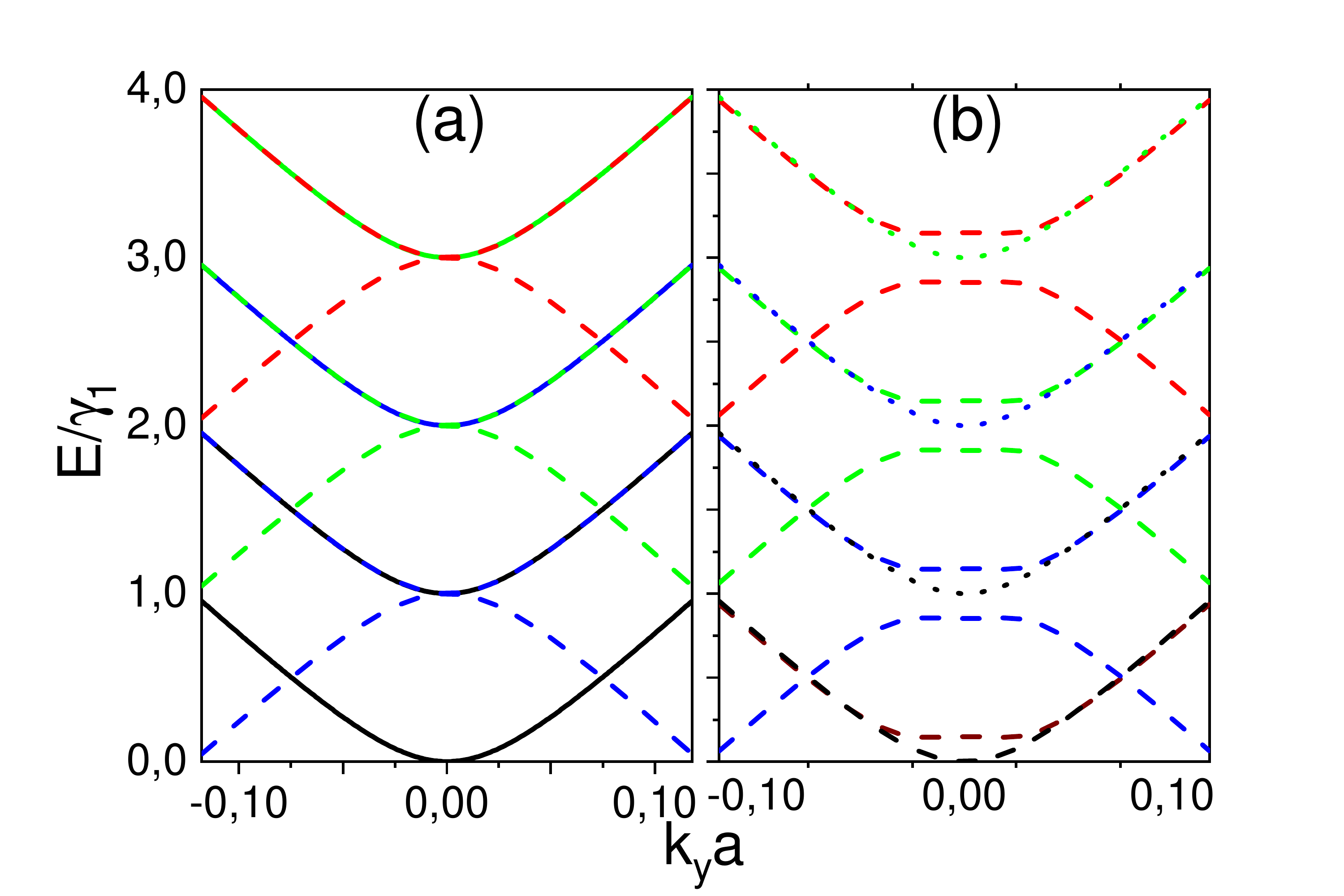}
	\caption{(Color online) The energy as a function of wave vector $k_y$. (a): Black lines represent the band's outside barriers while blue dashed line  ($U_{2}=1\gamma_{1}$), green dashed lines ($U_{3}=2\gamma_{1}$) and red dashed line  ($U_{4}=3\gamma_{1}$) show, respectively, the band's inside regions \text{2}, \text{3} and \text{4}. (b): The same as in (a), but with $\delta_2=\delta_3=\delta_4=0.15\gamma_1$, with the dotted lines represent the upper band in the conduction bands in each region.} 
	\label{fig2}
\end{figure}

 A general solution of the spinors can be determined in regions with potential barriers ($ j $ = 2, 3, 4) by solving the eigenvalue equation
 \begin{align}\label{eqva}
 H_ j\psi_ j=E_ j\psi_ j	
 \end{align} 
  and then we just have to set the potentials $V^{\pm}_{j}=0$ to obtain the solutions in both regions 1 and 5. By injecting Eqs. (\ref{E3}-\ref{E4}) into Eq. \eqref{eqva}, we end up with   four coupled differential equations. They are given by
\begin{subequations}
        \begin{align}
-i(\partial_{x}+k_{y})\phi^{j}_{B_{1}}&=(\varepsilon_{j}-\delta_{j})\phi^{j}_{A_{1}} \label{E5a}\\
-i(\partial_{x}-k_{y})\phi^{j}_{A_{1}}&=(\varepsilon_{j}-\delta_{j})\phi^{j}_{B_{1}}-\phi^{j}_{A_{2}}\label{E5b}\\
-i(\partial_{x}+k_{y})\phi^{j}_{B_{2}}&=(\varepsilon_{j}+\delta_{j})\phi^{j}_{A_{2}}-\phi^{j}_{B_{1}}\label{E5c}\\
-i(\partial_{x}-k_{y})\phi^{j}_{A_{2}}&=(\varepsilon_{j}+\delta_{j})\phi^{j}_{B_{2}}\label{E5d}
        \end{align}
\end{subequations}
where we have set $\varepsilon_{j}=E_{j}-U_{j}$, $k_{y}$ is the wave vector along $ y $-direction. 
By eliminating progressively the unknowns in the above equations, we obtain a differential equation of second degree. Choosing  $ \phi^{j}_{B_{1}} $,
for example, yields 
\begin{equation}
        \left[ \partial_{x}^{2}+(k^{\tau}_{j})^{2}\right] \phi^{j}_{B_{1}}=0,
        \label{E6}
\end{equation}
where $\tau=\pm$ indicates the propagation modes and $k^{\tau}_{j}$ is the corresponding longitudinal wave vector expressed as below 
\begin{equation}
k^{\tau}_{j}=\left[ \varepsilon_{j}^{2}+\delta_{j}^{2}-k_{y}^{2}+\tau\sqrt{\varepsilon_{j}^{2}(1+4\delta_{j}^{2})-\delta_{j}^{2}}  \right]^{1/2} 
\label{E7}
\end{equation}
\newpage
\noindent and then the energy is given by 
\begin{equation}
        \varepsilon^{\tau}_{\pm,j}=\pm\left[ k^{2}+\delta_{j}^{2} +\frac{1}{2}+\tau\sqrt{k^{2}(1+4\delta_{j}^{2})+\frac{1}{4}} \right]^{1/2}
        \label{E8}
\end{equation}
where the wave vector is $k=[{(k^{\tau}_{j})^{2}+k_{y}^{2}}]^{\frac{1}{2}}$. For  regions 1 and 5, where there is no potential barrier, the corresponding wave vector and energy spectrum can be derived from 
Eqs. (\ref{E7}-\ref{E8})  as follows
  \begin{align}
  &      k^{\tau}_{0}=[\varepsilon_{j}^{2}+\tau\varepsilon_{j}-k^{2}_{y}]^{1/2}
        \label{E9}\\
 & E^{\tau}_{\pm}=\pm\left[ k^{2}+\frac{1}{2}+\tau\sqrt{k^{2}+\frac{1}{4}} \right]^{1/2}  
  \label{E10}.      
  \end{align}
Fig. \ref{fig2} displays the energy spectrum of each region of the system presented in Fig. \ref{fig:png2pdf}. In the absence of an interlayer potential difference ($\delta_j =0$), we observe two bands that touch at $k=0$ as well as two others that are separated. 
As shown in Fig. \ref{fig2}(a), the result is that the upper band in the conduction band of one region coincides with the lower band in the conduction band of the adjacent region. 
When $\delta_j \neq 0$, the two touching bands separate, resulting in a gap, as shown in Fig.  \ref{fig2}(b). 
Hence, in the conduction band, the upper band (dotted lines) of a region and the lower (dashed lines) band of the next region are distinguished. If the wave vector in the first region $k^+_0$ is real while $k^-_0$ is imaginary, then propagation is only possible in the $k^+_0$ mode. when $E>\gamma_{1}$, $k^{-}_{0}$ is real too, propagation is possible via both $k^{+}_{0}$ and $k^{-}_{0}$ modes.

Now let us return to Eq. (\ref{E6}) and write its solution as a linear combination of plane waves. This is  
\begin{equation}
        \phi^{j}_{B_{1}}= c_{1}e^{i k^{+}_{j}x}+c_{2}e^{-i k^{+}_{j}x}+c_{3}e^{i k^{-}_{j}x}+c_{4}e^{-i k^{-}_{j}x}
        \label{E11}
\end{equation}
where $c_n$ denotes the normalization coefficients and $n=1,2,3,4$. One can easily derive the other components of the spinor by injecting (\ref{E11}) into (\ref{E5a}-\ref{E5d}). This process yields the following results
\begin{widetext}
\begin{align}
        \phi^{j}_{A_{1}}&= c_{1}f^{+}_{-}e^{i k^{+}_{j}x}-c_{2}f^{+}_{+}e^{-i k^{+}_{j}x}+c_{3}f^{-}_{-}e^{i k^{-}_{j}x}-c_{4}f^{-}_{+}e^{-i k^{-}_{j}x}
        \label{E12}\\
        \phi^{j}_{A_{2}}&= c_{1}h^{+}e^{i k^{+}_{j}x}+c_{2}h^{+}e^{-i k^{+}_{j}x}+c_{3}h^{-}e^{i k^{-}_{j}x}+c_{4}h^{-}e^{-i k^{-}_{j}x}
        \label{E13}\\
                \phi^{j}_{B_{2}}&= c_{1}h^{+}g^{+}_{+}e^{i k^{+}_{j}x}-c_{2}h^{+}g^{+}_{-}e^{-i k^{+}_{j}x}+c_{3}h^{-}g^{-}_{+}e^{i k^{-}_{j}x}-c_{4}h^{-}g^{-}_{-}e^{-i k^{-}_{j}x}
        \label{E14}
\end{align}
\end{widetext}
and the quantities
\begin{align}
      &  f^{\tau}_{\pm}=\frac{k^{\tau}_{j}\pm i k_{y}}{\varepsilon_{j}-\delta_{j}}
        \label{E15}
\\
& h^{\tau}=\frac{(\varepsilon_{j}-\delta_{j})^{2}-(k^{\tau}_{j})^{2}-k^{2}_{y}}{\varepsilon_{j}-\delta_{j}}
        \label{E16}
\\
&
        g^{\tau}_{\pm}=\frac{k^{\tau}_{j}\pm i k_{y}}{\varepsilon_{j}+\delta_{j}}
        \label{E17}
\end{align}
have been defined.
It is more suitable to write the general solution in matrix form in each region 
\begin{equation}
\psi^{j}(x,y)=G_{j} M_{j}(x) C_{j} e^{ik_{y}y}
\label{E18}
\end{equation} 
where $C_{j}$ represents the matrix form of coefficients $c_{n}$, and the matrix $G_{j}$ is given by 
\begin{equation}
        G_{j}=\begin{pmatrix}
        f^{+}_{-}& -f^{+}_{+} & f^{-}_{-}&-f^{-}_{+}\\
        1&1&1&1\\
        h^{+}&h^{+}&h^{-}&h^{-}\\
        h^{+}g^{+}_{+}&-h^{+}g^{+}_{-}&h^{-}g^{-}_{+}&-h^{-}g^{-}_{-}\\
        \end{pmatrix}
\label{E19}
\end{equation} \\
 as well as $M_{j}(x)$
\begin{equation}
        M_{j}(x)=\begin{pmatrix}
        e^{ik^{+}_{j}x}&0&0&0\\
        0&e^{-ik^{+}_{j}x}&0&0\\
        0&0&e^{ik^{-}_{j}x}&0\\
        0&0&0&e^{-ik^{-}_{j}x}\\
        \end{pmatrix}
\label{E20}.
\end{equation} 
As is obvious, solutions for regions outside the barriers are obtained by imposing $U_ j = \delta_j =0$. Accordingly, in region 1, we have 
\begin{widetext}
        \begin{align}
                \phi^{\text{1}}_{A_{1}}&= \delta_{\tau,1}f^{+}_{-}e^{i k^{+}_{0}x}-r^{\tau}_{+}f^{+}_{+}e^{-i k^{+}_{0}x}+\delta_{\tau,-1}f^{-}_{-}e^{i k^{-}_{0}x}-r^{\tau}_{-}f^{-}_{+}e^{-i k^{-}_{0}x}
                \label{E21}\\
                \phi^{\text{1}}_{B_{1}}&=\delta_{\tau,1}e^{i k^{+}_{0}x}+r^{\tau}_{+}e^{-i k^{+}_{0}x}+\delta_{\tau,-1}e^{i k^{-}_{0}x}+r^{\tau}_{-}e^{-i k^{-}_{0}x}
                \label{E22}\\
                \phi^{\text{1}}_{A_{2}}&= \delta_{\tau,1}h^{+}e^{i k^{+}_{0}x}+r^{\tau}_{+}h^{+}e^{-i k^{+}_{0}x}+\delta_{\tau,-1}h^{-}e^{i k^{-}_{0}x}+r^{\tau}_{-}h^{-}e^{-i k^{-}_{0}x}
                \label{E23}\\
                \phi^{\text{1}}_{B_{2}}&= \delta_{\tau,1}h^{+}g^{+}_{+}e^{i k^{+}_{0}x}-r^{\tau}_{+}h^{+}g^{+}_{-}e^{-i k^{+}_{0}x}+\delta_{\tau,-1}h^{-}g^{-}_{+}e^{i k^{-}_{0}x}-r^{\tau}_{-}h^{-}g^{-}_{-}e^{-i k^{-}_{0}x}
                \label{E24}
        \end{align}
\end{widetext}
and we write in region 5 
\begin{align}
\phi^{\text{5}}_{A_{1}}&= t^{\tau}_{+}f^{+}_{-}e^{i k^{+}_{0}x}+t^{\tau}_{-}f^{-}_{-}e^{i k^{-}_{0}x}
\label{E25}\\
\phi^{\text{5}}_{B_{1}}&= t^{\tau}_{+}e^{i k^{+}_{0}x}+t^{\tau}_{-}e^{i k^{-}_{0}x}
\label{E26}\\
\phi^{\text{5}}_{A_{2}}&= t^{\tau}_{+}h^{+}e^{i k^{+}_{0}x}+t^{\tau}_{-}h^{-}e^{i k^{-}_{0}x}
\label{E27}\\
\phi^{\text{5}}_{B_{2}}&= t^{\tau}_{+}h^{+}g^{+}_{+}e^{i k^{+}_{0}x}+t^{\tau}_{-}h^{-}g^{-}_{+}e^{i k^{-}_{0}x}
\label{E28}.
\end{align}

The next section will be devoted to the calculation of the transmission and reflection probabilities as well as the conductance. Given that there is no potential in the incident  ($j$=1) and the transmitted ($j$= 5) regions, we will use the following relation
\begin{equation}
G_{\text{1}} M_{\text{1}}(x)=G_{\text{5}}M_{\text{5}}(x)
\label{E29}.
\end{equation}

\section{Transport properties}\label{Trans}
In order to establish the transmission and reflection probabilities, it is necessary to impose the continuity of the spinors at the different interfaces of the triple barrier structure. In addition, by using the matrix transfer method, it becomes easy to connect the coefficients of the incident region to those of the transmitted one. These coefficients are given by 
\begin{equation}
        C_{\text{1}}^{\tau}=\begin{pmatrix}
                \delta_{\tau, 1} \\
                r_{+}^{\tau} \\
                \delta_{\tau,-1} \\
                r_{-}^{\tau}
        \end{pmatrix},\quad
        C_{\text{5}}^{\tau}=\begin{pmatrix}
                t_{+}^{\tau} \\
                0 \\
                t_{-}^{\tau} \\
                0
        \end{pmatrix}
\label{E30}
\end{equation}
where the Kronecker delta symbol is $\delta_{\tau,\pm1}$.
Using the continuity at interfaces $x=0, a, b, c$, the following results are obtained
\begin{align}
                &G_{\text 1} M_{\text 1}(0) C_{1}=G_{\text{2}} M_{\text{2}}(0) C_{\text{2}} \\
                \label{E31}
                &G_{\text{2}} M_{\text{2}}(a) C_{\text{2}}=G_{\text{3}} M_{\text{3}}(a) C_{\text{3}} \\
                \label{E32}
                &G_{\text{3}} M_{\text{3}}(b) C_{\text{3}}=G_{\text{4}} M_{\text{4}}(b) C_{\text{4}} \\
                &G_{\text{4}} M_{\text{4}}(c) C_{\text{4}}=G_{\text{5}} M_{\text{5}}(c) C_{\text{5}}
                \label{E34}.
        \end{align}
    As a result, using the transfer matrix $N$, the coefficients
 $C_{\text{1}}^{\tau}$ and $C_{\text{5}}^{\tau}$ can be connected
 as follows
\begin{equation}
        C_{\text{1}}^{\tau}=N C_{\text{5}}^{\tau}
        \label{E35}.
\end{equation}
where $N$ is defined as 
\begin{equation}
        N=\prod_{j=\text{1}}^{\text{4}} M_{j}^{-1}(x_{j}) G_{j}^{-1} G_{j+1} M_{j+1}(x_{j})
        \label{E36}.
\end{equation}
Therefore, we can find the transmission and reflection coefficients from
\begin{equation}
        \begin{pmatrix}
                t_{+}^{\tau} \\
                r_{+}^{\tau} \\
                t_{-}^{\tau} \\
                r_{-}^{\tau} 
        \end{pmatrix}=\begin{pmatrix} 
                N_{11} & 0 & N_{13} & 0 \\
                N_{21} & -1 & N_{23} & 0 \\
                N_{31} & 0 & N_{33} & 0 \\
                N_{41} & 0 & N_{43} & -1
            \end{pmatrix}^{-1} \begin{pmatrix} 
                \delta_{\tau, 1} \\
                0 \\
                \delta_{\tau,-1} \\
                0
        \end{pmatrix}
\label{E37}
\end{equation}
where $N_{ij}$ are the matrix elements of  $N$ given in Eq. \eqref{E36}. Finally, we end up with the transmission and reflection coefficients 
\begin{align}
&t_{+}^{\tau}=\frac{N_{33} \delta_{\tau, 1}-N_{13} \delta_{\tau,-1}}{N_{11} N_{33}-N_{13} N_{31}}, \quad t_{-}^{\tau}=\frac{N_{11} \delta_{\tau,-1}-N_{31} \delta_{\tau, 1}}{N_{11} N_{33}-N_{13} N_{31}} 
\label{E38}
\\
&
r_{+}^{\tau}=N_{21} t_{+}^{\tau}+N_{23} t_{-}^{\tau} 
,\quad
r_{-}^{\tau}=N_{41} t_{+}^{\tau}+N_{43} t_{-}^{\tau} 
\label{E39}.
\end{align}
To obtain the transmission and reflection probabilities, we use the current density. It is important to take into account the change in the wave velocity since the electrons can be scattered into different modes of propagation. The current density is given by 
\begin{equation}
\textbf{j}=v_{F}{\Psi}^{\dagger}\vec{\alpha}\Psi
\label{E42}
\end{equation}
where $\vec{\alpha}$ is a $4\times4$ matrix with two Pauli matrices $\sigma_{x}$ on its  diagonal and the other elements are zero.
We can now derive 
the incident $\textbf{j}_{\text{inc}}$, reflected $\textbf{j}_{\text{ref}}$ and transmitted $\textbf{j}_{\text{tra}}$ current densities from Eq. \eqref{E42}.
As a result, the transmission and reflection probabilities are given by
\begin{align}
   &     T_{\pm}^{\tau}=\frac{\left|\textbf{j}_{\text{tra}}\right|}{\left|\textbf{j}_{\text{inc}}\right|}=\frac{k_{0}^{\pm}}{k_{0}^{\tau}}\left|t_{\pm}^{\tau}\right|^{2}
        \label{E43}
\\
&
 R_{\pm}^{\tau}=\frac{\left|\textbf{j}_{\text{ref}}\right|}{\left|\textbf{j}_{\text{inc}}\right|}=\frac{k_{0}^{\pm}}{k_{0}^{\tau}}\left|r_{\pm}^{\tau}\right|^{2}
                \label{E44}.
\end{align}
The conservation of the total probability results in
\begin{equation}
        \sum_{i,j}\left(T_{i}^{j}+R_{i}^{j}\right)=1
        \label{E45}
\end{equation}
where $i,j=\pm$. 
The index $i$ points to the arriving mode, while the index $j$ points to the exiting mode. Consider the channel $k^+$, for example, and we get $ T_{+}^{+}+T_{+}^{-}+R_{+}^{+}+R_{+}^{-}=1. $

Since the transmission probabilities have already been determined, one can calculate the conductance according to the formula of Landauer-Büttiker \cite{BLANTER20001}
\begin{equation}
        \mathrm{G}(E)=G_{0} \frac{L_{y}}{2 \pi} \int_{-\infty}^{+\infty} \mathrm{d} k_{y} \sum_{\tau, n=\pm} T_{n}^{\tau}\left(E, k_{y}\right)
\end{equation}
with $L_{y}$ the width of the sample in the $ y $-direction and $G_{0}=4 \frac{\mathrm{e}^{2}}{h}$, the factor 4 is the result of the valley and spin degeneracy in graphene.

\section{Results and discussions}\label{RRDD}
\subsection{Two band Tunneling}

\begin{figure}\centering
        \includegraphics[width=0.5\textwidth]{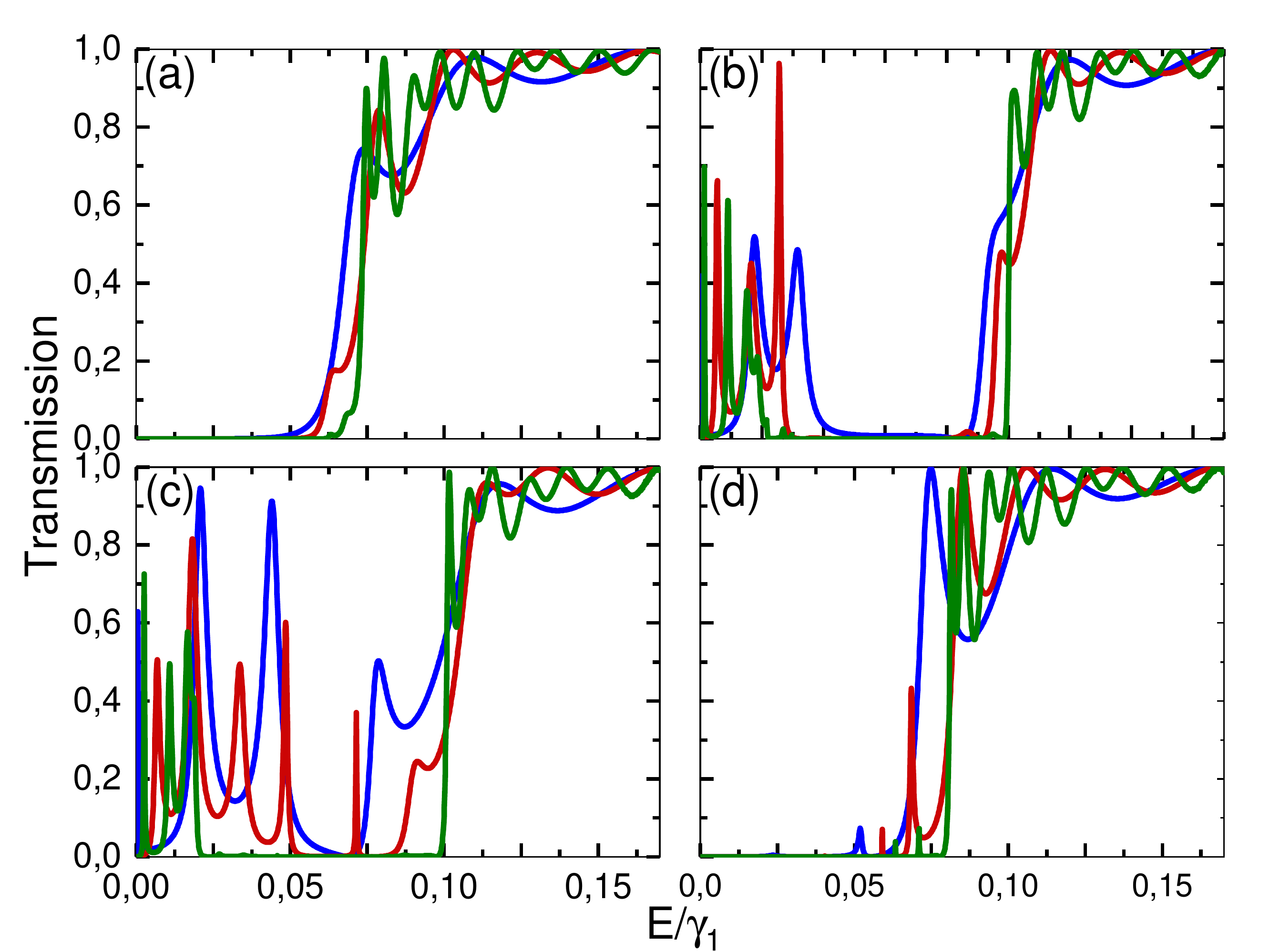}
       
        \caption{(Color online): Transmission  as a function of  energy $ E $, at normal incidence ($ k_y=0 $), for barrier widths $b_{1}=b_{2}=b_{3}= 15$ nm (blue line), $25$ nm (red line), $60$ (green line) and $U_{2}=0.05\gamma_{1}$, $U_{3}=0.06\gamma_{1}$, $U_{4}=0.07\gamma_{1}$. (a): $\delta_{2}=\delta_{3}=\delta_{4}=0$. (b): $\delta_{2}=\delta_{3}=\delta_{4}=0.03\gamma_{1}$. (c): $\delta_{2}=\delta_{4}=0.03\gamma_{1}$ and $\delta_{3}=0$. (d): $\delta_{2}=0.03\gamma_{1}$ and $\delta_{3}=\delta_{4}=0$. }
        \label{fig3}
\end{figure}

As mentioned in section \ref{TTMM}, propagation is via a single mode in the low energy regime ($E<\gamma_{1}$), so there is only one transmission channel. In Fig. \ref{fig3}, we show the transmission  as a function of  energy $ E $ at normal incidence ($k_y=0$) for various barrier widths:  $b_ 1=b_ 2=b_ 3=15$ nm (blue line),  $25$ nm (red line), and  $60$ nm (green line). Fig. \ref{fig3}(a) shows the transmission for barriers of different heights $U_{2}=0.05\gamma_{1}$, $U_{3}=0.06\gamma_{1}$, $U_{4}=0.07\gamma_{1}$ and $\delta_{2}=\delta_{3}=\delta_{4}=0$.
 The transmission behaves analogously to the one obtained for a single barrier ($d=100$ nm) in \cite{Van}, even for small barrier widths (blue and red lines). In fact, in the first energy range, $E<U_{2}$, we observe the manifestation of anti-Klein tunneling, which means that the transmission is zero. This is due to the chiral nature of charge carriers in graphene \cite{Katsnelson_2006}. However, for energy greater than $U_{2}$ transmission resonances appear. When an interlayer bias is introduced, the transmission probability resembles that of a single barrier ($d=100$ nm) in gapped bilayer graphene (see \cite{BENLAKHOUY2021114835}), but without resonance peaks in the energy range
 $E<U_{2}-\delta$. 
In addition, at the edge of the induced gap, transmission appears for non-large barrier widths, as observed in the case of double barrier \cite{Mouhafid}. Nevertheless, when we significantly increase the barrier width (green line), no transmission appears in the gap region. We conclude that the gap size strongly depends  on the barrier widths. In Fig. \ref{fig3}(c), we set all parameters except $\delta_3$ to 0, as in Fig. \ref{fig3}(b). For a large barrier width (green line), it is evident that the gap region stays the same even though $\delta_{3}$ is zero, while for a small value, the gap region narrows in favor of transmission. Once again, we see the gap can be adjusted by the barrier widths. In Fig. \ref{fig3} (d), with $\delta_{2}=0.03\gamma_{1}$ and $\delta_{3}=\delta_{4}=0$, the transmission  presents a similar aspect to that obtained in Fig. \ref{fig3}(a). The bias introduced in region 2 of the triple barrier system cannot open a gap on its own, even for large values of the barrier widths.

\subsection{Four band Tunneling} 
\begin{figure}
        \centering
        \includegraphics[width=0.54\textwidth]{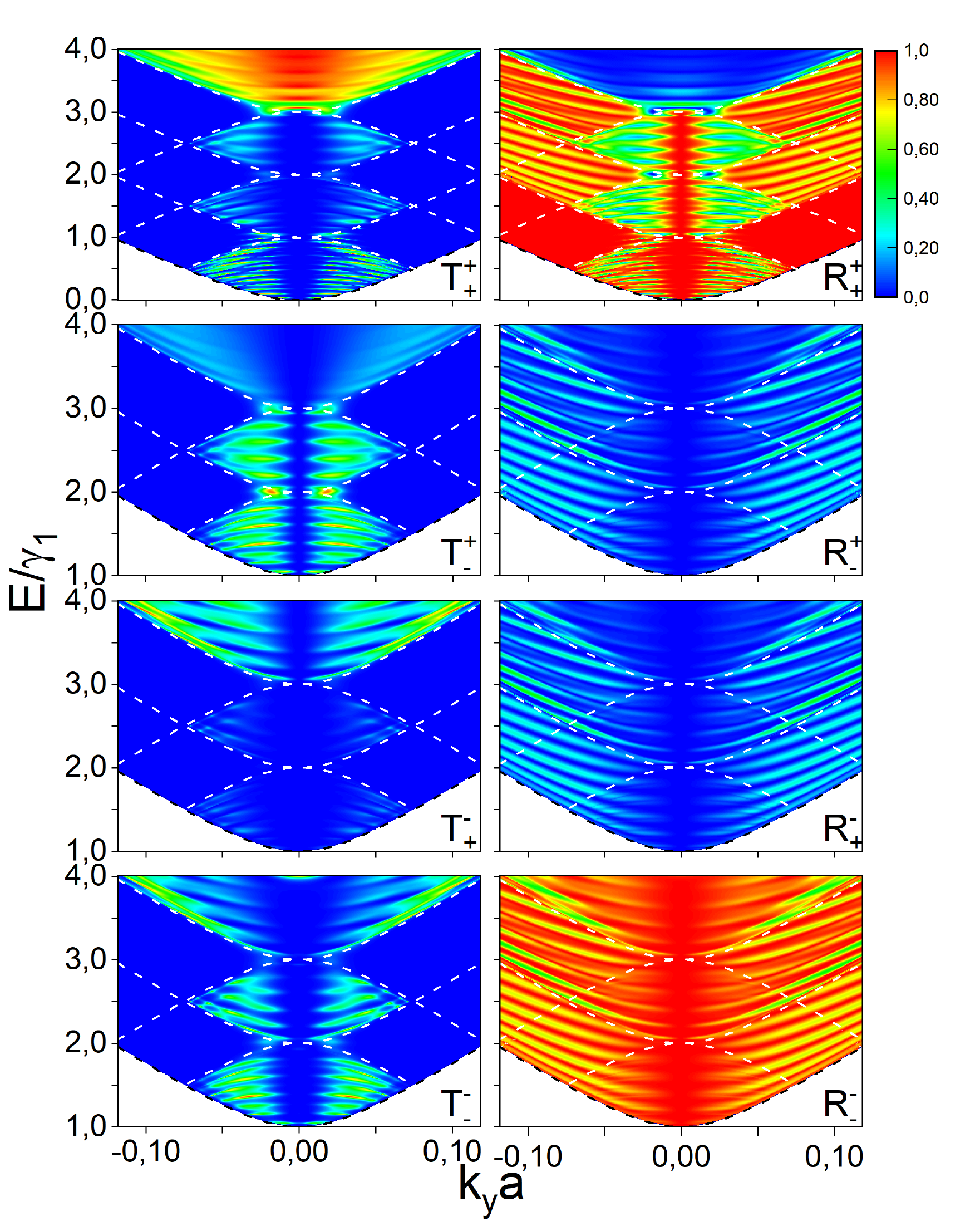}
        \caption{(Color online): Density plot of transmission and reflection  as a function of energy $ E $ and the wave vector $k_{y}$ for barrier heights $U_{2}=1\gamma_{1}$, $U_{3}=2 \gamma_{1}$, $U_{4}=3 \gamma_{1}$ and $\delta_{2}=\delta_{3}=\delta_{4}=0$. The barrier widths are set to be the same, $b_{1}=b_{2}=b_{3}=20$ nm. The dashed black lines represent the band's outside barriers while the dashed white lines show the band's inside barriers.}
        \label{fig:tandrdelta0}
\end{figure}

For higher energy $(E>\gamma_{1})$, the availability of two propagation modes gives rise to four transmission and four reflection channels, which we plot as a function of energy $ E $ and transverse wave vector $k_{y}$ in Fig. \ref{fig:tandrdelta0}. We fixed the barrier heights at $U_ 2 = \gamma_ 1 $, $U_ 3 = 2 \gamma_ 1 $, and $U_ 4 = 3 \gamma_ 1 $, as well as the barrier width at $b_ 1 =b_ 2 =b_ 4 =20$ nm and the interlayer potential difference at $\delta_ 3=\delta_ 4=0$. 
As shown in Fig. \ref{fig:tandrdelta0}, there are different energy regions delimited by dashed lines, which represent the propagating modes inside and outside the barriers.
The cloak effect occurs in the region of energy beneath $U_2$ at nearly normal incidence where the two modes $k^+$ outside and $k^-$ inside are decoupled, resulting in no scattering between them. In the same region, for non-normal incidence, the two modes $k^{+}$ outside and $k^{-}$ inside are coupled, then transmission appears. However, unlike the case of unbiased single barrier \cite{Van}, no resonance occurs. For nearly normal incidence, the cloak effect appears in the energy regions $U_{2}<E<U_{3}$ and $U_{3}<E<U_{4}$. 
The asymmetry of the triple barrier system breaks the equivalence observed between $T_-^+$ and $T_+^-$ in the case of unbiased single barrier \cite{Van}. It is clear that the cloak effect occurs exclusively for nearly normal incidence in the $T_-^+$ channel and for energy regions $U_{2}<E<U_{3}$ and $U_{3}<E<U_{4}$, as few transmission resonances appear.  
On the other hand, in the $T_+^-$ channel, the cloaking persists even for non-normal incidence so that there is no transmission in the same energy regions. However, the reflection channels $R_+^-$ and $R_-^+$ remain equivalent because the incident electrons return to the electron state \cite{Van,Mouhafid}. Transmission is suppressed in the $T_-^-$ channel in the two energy ranges of
$U_{2}<E<U_{3}$ and $U_{3}<E<U_{4}$, as in single and double barrier cases \cite{Van,BENLAKHOUY2021114835,Mouhafid}. 

\begin{figure}
        \centering
        \includegraphics[width=0.54\textwidth]{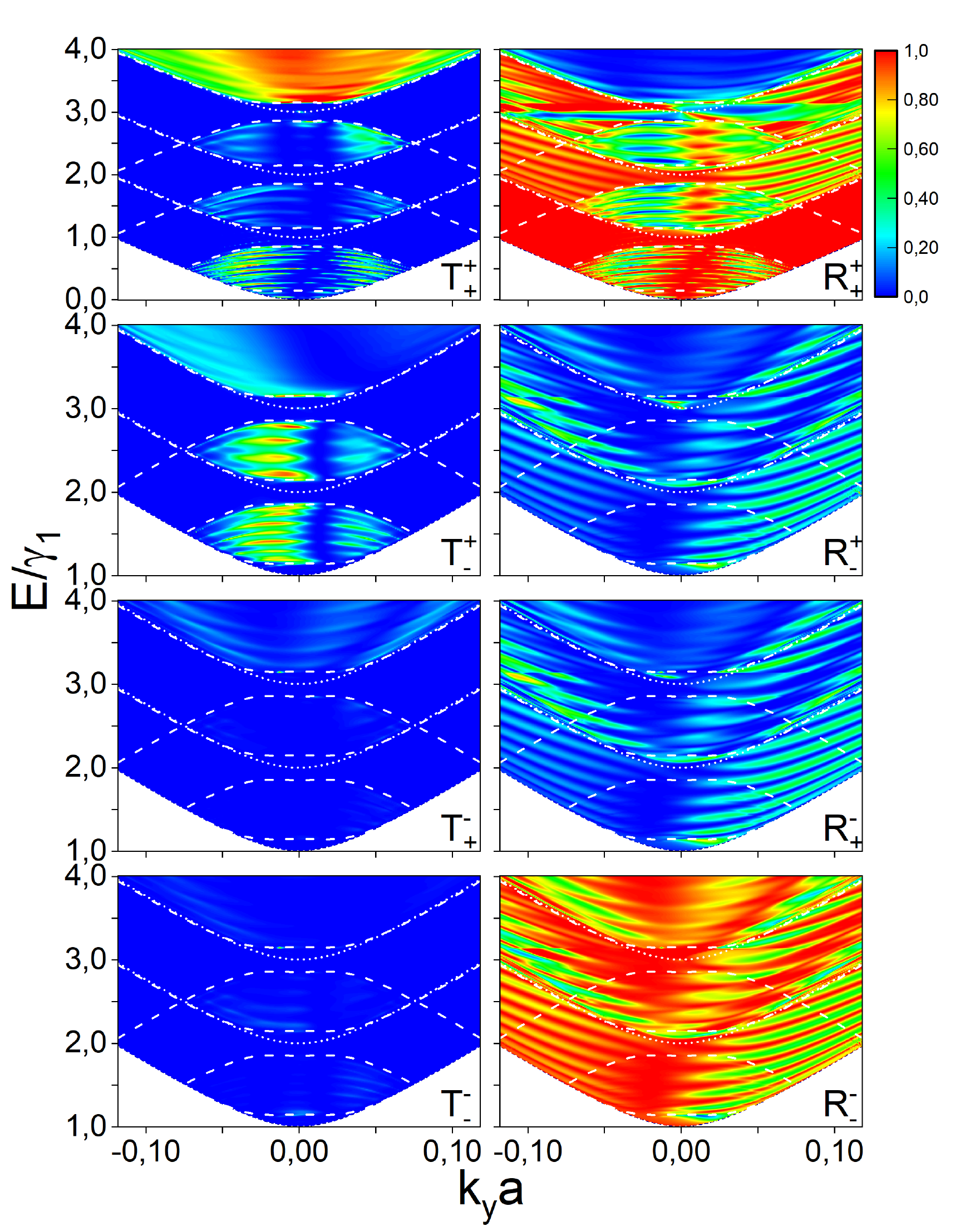}
                \caption{(Color online) The same as in Fig. \ref{fig:tandrdelta0}, but now for $\delta_{2}=\delta_{3}=\delta_{4}=0.15$. The dotted black lines represent the band's outside barriers while the dashed and dotted white lines show the band's inside barriers.}
        \label{fig:tandrdelta015}
\end{figure}

To assess the impact of the interlayer potential difference, we plot the transmission and reflection  as a function of energy $ E $ and wave vector $k_y$ in Fig. \ref{fig:tandrdelta015} using the same parameters as in Fig. \ref{fig:tandrdelta0}, but with $\delta_2=\delta_3=\delta_4=0.15 \gamma_1$. 
In the cases of double barriers, without \cite{Mouhafid} and with a magnetic field, \cite{Redouani}, transmissions have occurred in the gap as a result of the bound states in the well. Because there is no well in the triple barrier system, transmission is completely suppressed in the gap regions $U_ j-\delta_j<E<U_ j+\delta_j$, $ j $=2, 3, 4. 
However, despite the absence of a well, for small barrier widths, transmission occurs. We notice that the more we increase the width of the barriers, the more the transmission is suppressed in the gap regions. The cloak effect is slightly modified in the $T_{+}^{+}$ channel and for the energy range $E<U_2$, and thin transmission resonances appear, but they are smaller than those observed in \cite{Van}. 
The $T_{-}^{+}$  channel exhibits more transmission resonances than the result shown in Fig. \ref{fig:tandrdelta0}. In the presence of bias, no resonance appears in the $T_{-}^{+}$ channel in the case of single \cite{Van} and double barriers \cite{Mouhafid}, in contrast to our case. In energy regions $U_{2}<E<U_{3}$ and $U_{3}<E<U_{4}$, however, there is no transmission in $T_{+}^{-}$ and $T_{-}^{-}$   channels.

\begin{figure*}
        \centering
      \includegraphics[width=1\textwidth]{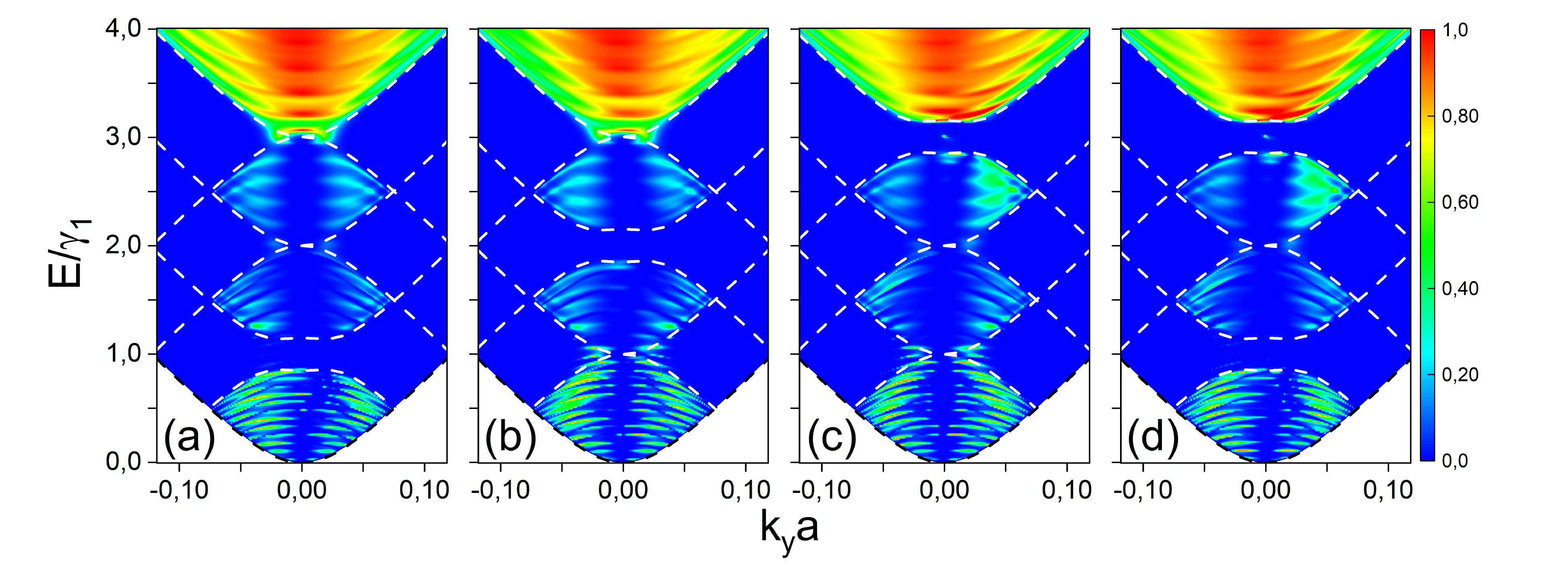}
        \caption{(Color online) Density plot of transmission  $T_{+}^{+}$ as a function of energy $ E $ and the wave vector $k_{y}$ for barrier heights $U_{2}=1 \gamma_{1}$, $U_{3}=2 \gamma_{1}$, $U_{4}=3 \gamma_{1}$. (a): $\delta_{2}=0.15\gamma_{1}$,  $\delta_{3}=\delta_{4}=0.$ (b): $\delta_{3}=0.15\gamma_{1}$,  $\delta_{2}=\delta_{4}=0.$ (c): $\delta_{4}=0.15\gamma_{1}$,  $\delta_{2}=\delta_{3}=0.$ (d): $\delta_{3}=0$,  $\delta_{2}=\delta_{4}=0.15\gamma_{1}.$ The barrier widths are set to be the same, $b_{1}=b_{2}=b_{3}=20$ nm. The dashed black lines represent the band's outside barriers while the dashed white lines show the band's inside barriers.}
        \label{fig:Transdeltavariable}
\end{figure*}

The transmission probability $T_{+}^{+}$ is presented in 
Fig. \ref{fig:Transdeltavariable} 
with the same parameters as in Fig. \ref{fig:tandrdelta015} but for variable interlayer potential in each barrier. We chose $\delta_{2}=0.15\gamma_{1}$ and  $\delta_{3}=\delta_{4}=0.$ in Fig. \ref{fig:Transdeltavariable}(a),  $\delta_{3}=0.15\gamma_{1}$ and $\delta_{2}=\delta_{4}=0.$ in Fig. \ref{fig:Transdeltavariable}(b), $\delta_{4}=0.15\gamma_{1}$ and $\delta_{2}=\delta_{3}=0.$ in Fig. \ref{fig:Transdeltavariable}(c), finally $\delta_{3}=0$ and $\delta_{2}=\delta_{4}=0.15\gamma_{1}.$ in Fig. \ref{fig:Transdeltavariable}(d). 
Introducing bias in one of the triple barrier regions, in contrast to the low energy regime ($E<\gamma_1$), opens a gap. Furthermore, we see that the cloak effect is more adjusted, so the number of thin resonances in the energy zone under $U_2$ has increased in comparison to the $T_{+}^{+}$ channel in Fig. \ref{fig:tandrdelta015}. It is also worth noting that Figs. \ref{fig:Transdeltavariable}(a,d) have fewer resonances than Figs. \ref{fig:Transdeltavariable}(b,c), which is due to the gap induced in the region $U_ 2-\delta_2<E<U_ 2+\delta_2$. 


\subsection{Conductance}
	\begin{figure*}
\centering
  \includegraphics[width=0.32\textwidth]{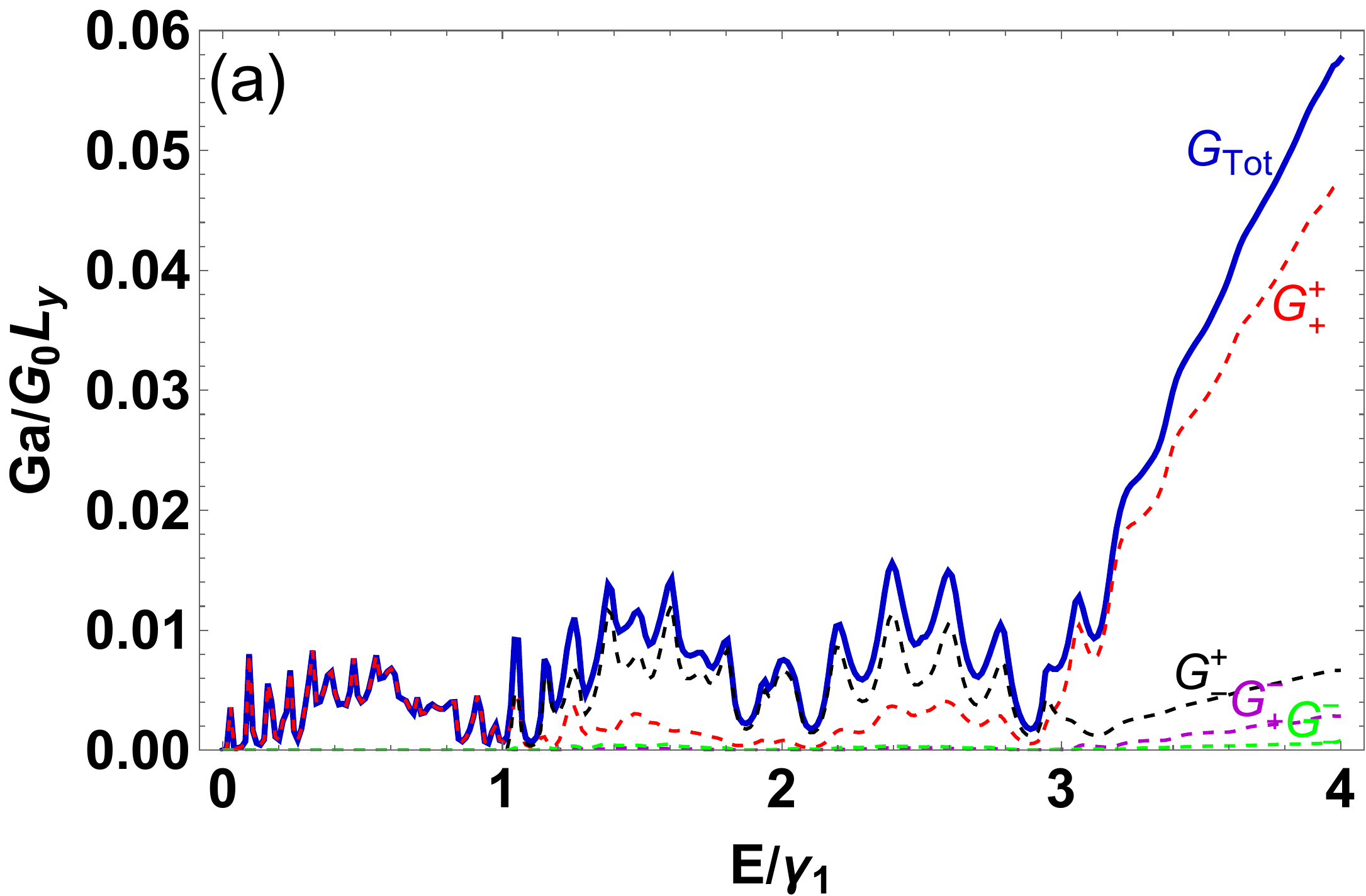}
  \includegraphics[width=0.32\textwidth]{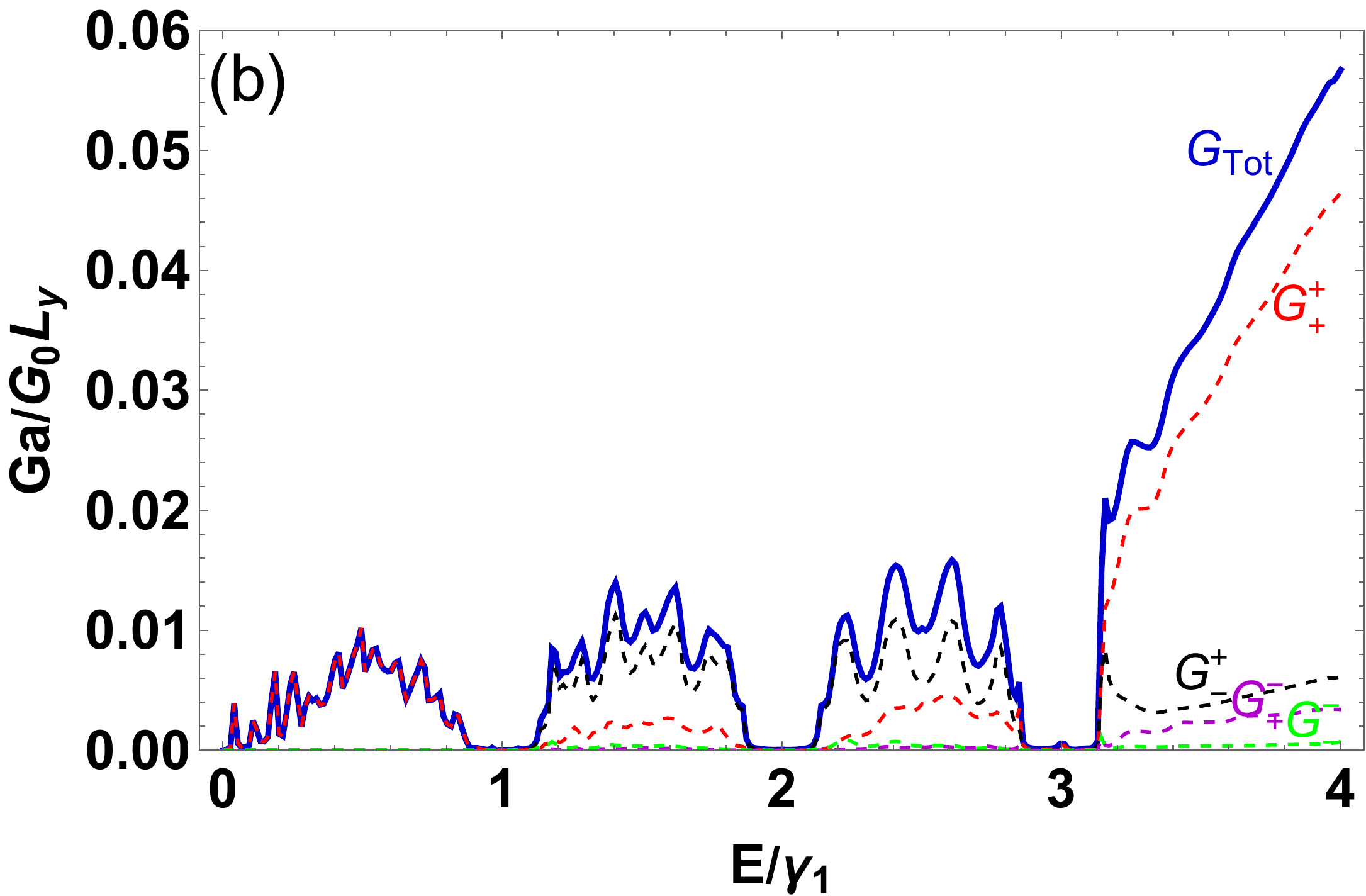}
  \includegraphics[width=0.32\textwidth]{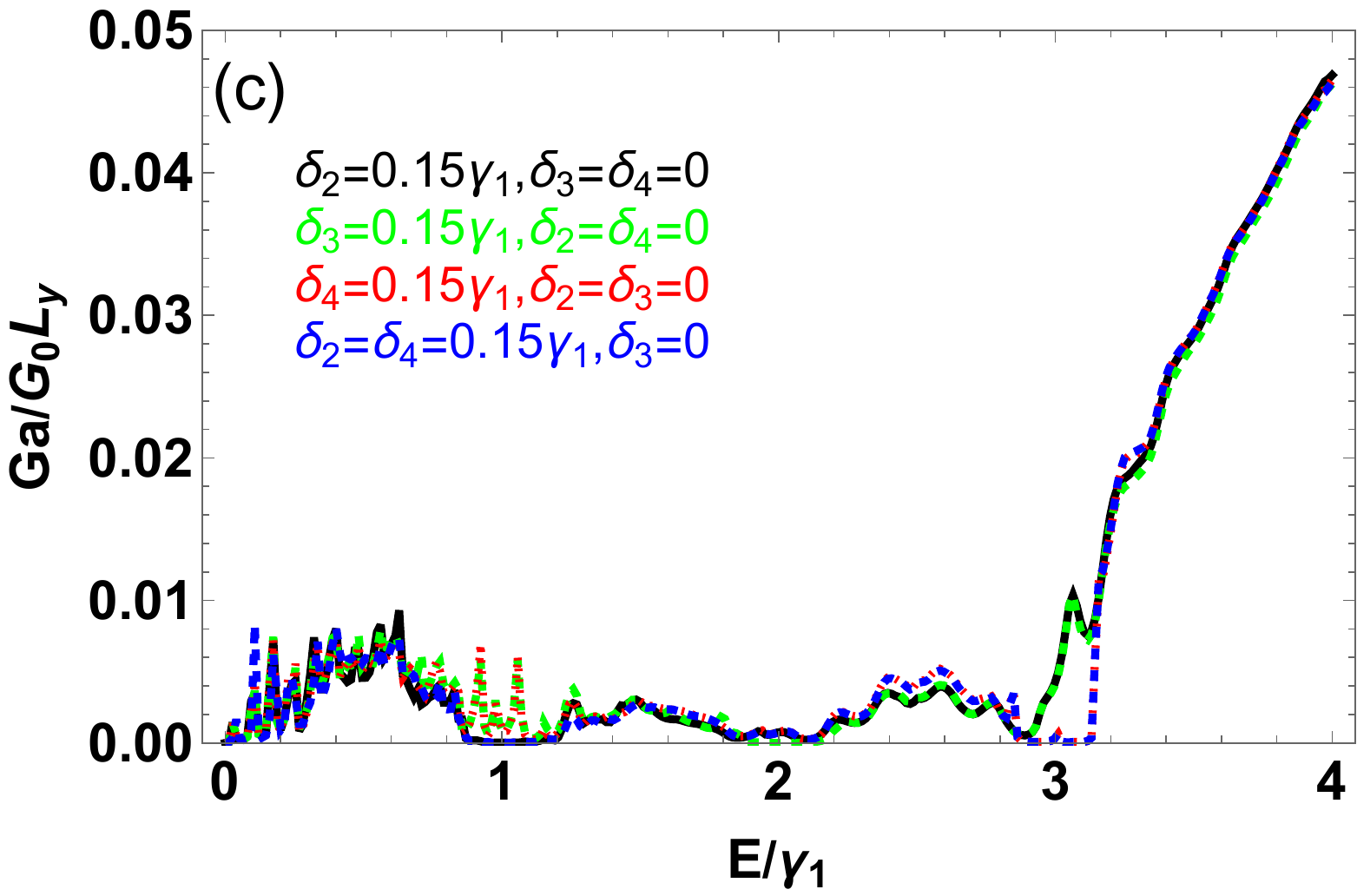}
       \caption{(Color online) Conductance of the triple barrier structure as a function of energy $ E $ for $U_{2}=1\gamma_{1}$, $U_{3}=2\gamma_{1}$, $U_{4}=3\gamma_{1}$ and $b_{1}=b_{2}=b_{3}=20$ nm. (a):  $\delta_{2}=\delta_{3}=\delta_{4}=0$. (b): $\delta_{2}=\delta_{3}=\delta_{4}=0.15 \gamma_{1}$. (c):  variable interlayer potential difference.}
\label{Conductancedelta}
\end{figure*}

In Fig. \ref{Conductancedelta}, we investigate the energy dependence of the conductance  for barrier heights $U_{2}=1\gamma_{1}$, $U_{3}=2\gamma_{1}$, $U_{4}=3\gamma_{1}$ and barrier widths $b_{1}=b_{2}=b_{3}=20$ nm. Fig. \ref{Conductancedelta}(a) shows the conductance in absence of interlayer potential difference, i.e. $\delta_{2}=\delta_{3}=\delta_{4}=0$. 
The resonances observed in the transmission probabilities (Fig. \ref{fig:tandrdelta0}) result in peaks in the triple barrier conductance. For energies less than $U_ 2$, the $T_{+}^{+}$  channel is the only one contributing to total conductance, i.e. $G_{Tot}=G_{+}^{+}$.   
In the energy region smaller than $U_{2}$, it is clear that, as a consequence of the contribution of $T_{-}^{+}$ shoulders appear in the total conductance, which increased. This is because resonances in transmission $T_{-}^{+}$ exist in this energy region.  
Also because of the cloaking, the contributions of $T_{+}^{-}$ and $T_{-}^{-}$ remain zero. However, for energies larger than $U_{4}$, $G_{+}^{-}$ starts conducting while $G_{-}^{-}$ stays close to zero. The contributions of $T_{-}^{+}$ and $T_{+}^{-}$ are equal in the case of simple barrier \cite{Van,BENLAKHOUY2021114835}, implying that $G_{-}^{+}=G_{+}^{-}$. In our case, the asymmetric aspect of the triple barrier breaks this equality, i.e. $G_{-}^{+}\ne G_{+}^{-}$.
The conductance for non-null values of interlayer potential difference, $\delta_{2}=\delta_{3}=\delta_{4}=0.15\gamma_{1}$, is plotted in Fig. \ref{Conductancedelta}(b). We see suppression of conductance in the gap regions $U_{j}-\delta_{j}<E<U_{j}+\delta_{j}$ ($ j $ = 2, 3, 4), resulting in a reduction in the number of shoulders. 
Fig. \ref{Conductancedelta}(c) shows the conductances $G_{+}^{+}$ associated with  transmission $T_{+}^{+}$
plotted in Fig. \ref{fig:Transdeltavariable}.
When $\delta_{2}=0.15\gamma_{1}$ (black and blue dashed lines) and $E<U_{2}+\delta_{2}$ are used,  $G_{+}^{+}$ has fewer peaks than when $\delta_ 2 = 0$ (green dashed and red dashed lines). 
 This is a consequence of the gap opened by $\delta_{2}=0.15\gamma_{1}$ around $E=U_{2}$. 
All of the conductances $G_{+}^{+}$ are roughly equivalent in the energy range $U_2+\delta_2$. It is worth noting that the conductance nears zero around $E=U_3$, even when $\delta_3=0$. To sum up, the main differences between the different cases are found around $E=U_{2}$ and $E=U_{4}$.

\section{Conclusion}\label{CC}

We have investigated the transport properties of AB bilayer graphene through a triple barrier system. Starting from the four band Hamiltonian, we have determined the energy spectrum and the eigenvectors of each region. Thereafter, based on the transfer matrix method and using the continuity conditions at the different interfaces, we have calculated the transmission and reflection probabilities. When the barrier width is reduced in the low energy regime ($E<\gamma_1$), at normal incidence ($ k_y=0 $), and in the presence of bias, it was seen that the number of transmission's peaks increase in the gap region. We have also observed that the application of bias in only one region is not enough to create a gap, even when the   barrier width is increased. However, the application of bias in two regions resulted in an opening  gap for high barrier widths.

Subsequently, we have evaluated the transmission for an energy regime greater than $\gamma_{1}$.  As a result, we have shown the appearance of extra resonances, especially in the $T_{-}^{+}$ transmission channel. Consequently, we have reached the following conclusions. Firstly, the application of the interlayer potential difference controls the number of gaps created. Secondly, the number of peaks in the conduction is increased compared to the case of the double barrier \cite{Mouhafid} because of the resonances present in $T_{-}^{+}$.

\end{document}